\documentclass[aip,jcp,groupedaddress,amsmath,amssymb,reprint]{revtex4-1}

\usepackage{graphicx}
\usepackage{dcolumn}
\usepackage{bm}
\usepackage{subfigure,xcolor}

\begin{document}


\title{Probing exciton/exciton interactions with entangled photons:theory}

\author{Eric R. Bittner}
\affiliation{Department of Chemistry \& Department of Physics, University of Houston, Houston, TX 77204}
\affiliation{Department of Physics, Durham University,
Durham, United Kingdom.
}
\email{Corresponding Author:ebittner@central.uh.edu}

\author{Hao Li}
\affiliation{Department of Chemistry, University of Houston, Houston, TX 77204}

\author{Andrei Piryatinski}
\affiliation{Theoretical Division, Los Alamos National Lab, Los Alamos, NM 87545}

\author{Ajay~Ram~Srimath~Kandada}
\affiliation{School of Chemistry \& Biochemistry and School of Physics, Georgia Institute of Technology, 901 Atlantic Drive, Atlanta, Georgia 30332}
\affiliation{Center for Nano Science and Technology @Polimi, Istituto Italiano di Tecnologia, via Giovanni Pascoli 70/3, 20133 Milano, Italy}

\author{Carlos Silva}
\affiliation{School of Chemistry \& Biochemistry and School of Physics, Georgia Institute of Technology, 901 Atlantic Drive, Atlanta, Georgia 30332}

\date{\today}

\begin{abstract}
Quantum entangled photons provide a sensitive probe
of many-body interactions and offer an unique
experimental portal 
for quantifying many-body correlations in a material system.
In this paper, we present a theoretical demonstration of
how photon-photon entanglement can be generated 
via interactions between coupled qubits.  Here
we develop a model for the scattering of an entangled pair 
of photons from a molecular dimer. We develop a
diagrammatic theory for the scattering matrix and show that one can 
correlate the von Neumann entropy of the outgoing bi-photon 
wave function to exciton exchange and repulsion interactions.
We conclude by discussing possible experimental scenarios 
for realizing these ideas.
\end{abstract}

\maketitle

\section{Introduction}

  As evidenced by recent 
theoretical and experimental 
advances,  quantum entangled
photons provide a sensitive measure
of collective and 
many-body dynamics.
 \cite{Ladd2010,Lemos2014,RevModPhys.74.145,Sewell2017,Zou1991a,PhysRevA.79.033832,Schlawin2013,Kalashnikov:PRX2014,Kalashnikov:SciRep2017,Mukamel:PRA2012,Goodson:JPCL2017,Villabona-Monsalve:JPCA2017,Mukamel:RevModPhys2016,AHMarcus:JPCB2013,MCTiech:PRL1998,Goodson:JPCL2013,Goodson:JACS2010,Goodson:JPCB2006,Yabushita:PRA2004,HaoLi:2018,Li:2019}
The sensitivity stems from the
``spooky action at a distance'' nature of 
 entangled photons, whereby measurement of one photon
gives information about its entangled partner photon.
This information 
can be extracted 
through either coincidence detection, interference, 
or quantum state reconstruction.   

We recently presented a theoretical basis for how 
entanglement can be produced in 2-photon scattering
from a system of coupled excitonic sites.\cite{HaoLi:2018,Li:2019}  In our 
approach, we assume that the bi-photon 
scattering matrix can be decomposed into 
a product of two single-photon terms and an irreducible
two-photon term of the form
\begin{eqnarray}
S^{(2)}(\omega_1,\omega_2;\omega_1',\omega_2')
&=&
S^{(1)}(\omega_1,\omega_1')
S^{(1)}(\omega_2,\omega_2') \nonumber \\
&\times& e^{g(\omega_1,\omega_2;\omega_1',\omega_2')}
\end{eqnarray}
where $S^{(1)}$ gives the 
single photon (Raman or Rayleigh) scattering and
$g(\omega_1,\omega_2;\omega_1',\omega_2')$
is an irreducible term that can be related
to exciton/exciton cross-correlations.\cite{Li:2019} 
We suggest that by measuring the photon
entanglement entropy change, one can deduce a direct measure
of exciton/exciton cross-correlations. 
Here, we perform a theoretical analysis of the
two-photon scattering produced by a simple two-qubit 
system, coupled by exchange interactions which allow 
a single excitation to be transferred between qubits
and a direct interaction which introduces an energetic 
cost for double excitation.  
We show that the bi-photon scattering can be 
related to the cascade emission from the double-excited
system, can be ``tuned'' by 
changing the nature of the exchange term. 

Significant amounts of theoretical and experimental efforts have been invested to achieve photon pair {\em polarization} entanglement using photon cascades in semiconductor quantum dots.\cite{Moreau_PRL:2001,Akopian_PRL:2006,Avron_PRL:2008,MullerNatPhot:2014,HuberJOpt:2018} In that case, special conditions making indistinguishable two alternative emission passes via split intermediate states need to be satisfied.\cite{Akopian_PRL:2006,Avron_PRL:2008} In contrast, below we study the {\em energy/time} photon entanglement generation which turns out to have much less restrictions to be achieved.

\section{Theoretical Model}
\begin{figure*}[th]
    \centering
\subfigure[]{\includegraphics[width=0.95\columnwidth]{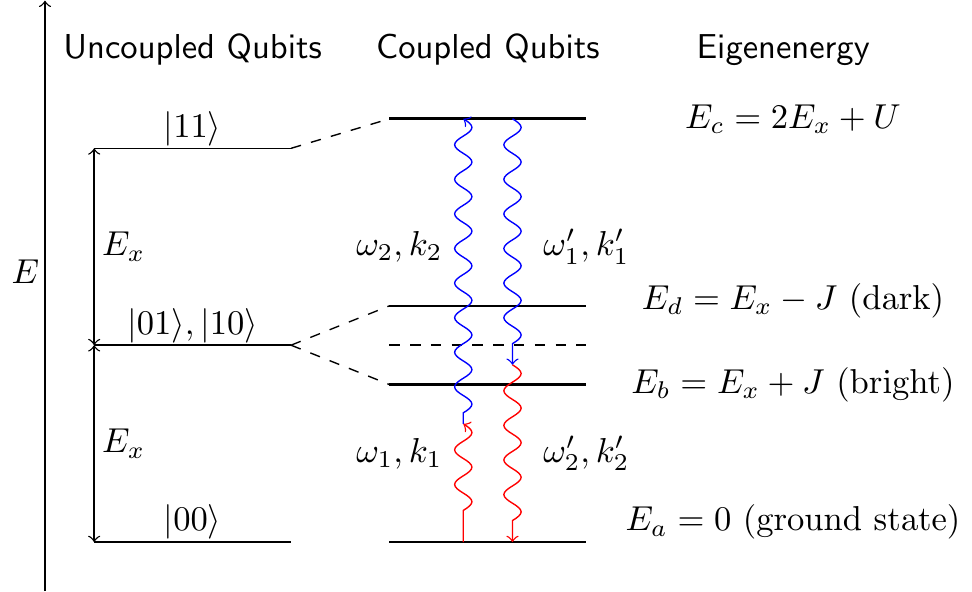}}
\subfigure[]{\includegraphics[width=0.95\columnwidth]{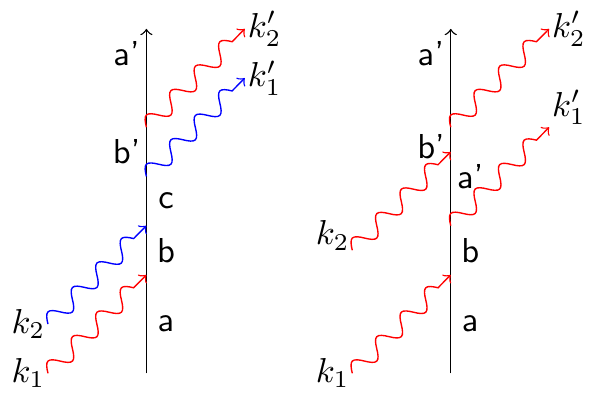}}
    \caption{(a) Energy Level Diagram for pair
    of interacting excitons. 
    In our model system, we assume that $J<0$ and $U>0$. 
    To the left are the uncoupled energy levels.  The exchange $J$ and anharmonicity $U$ mixes the $|10\rangle$, $|01\rangle$ and $|11\rangle$ configuration; however, one of these $(|10\rangle-|01\rangle)/\sqrt{2}$ with energy $E_x-J$ does not couple to the radiation field within the dipole/rotating wave approximation. 
    Physically, this 
    corresponds to a molecular dimer forming a 
    $J$-aggregate.   
    (b)Feynman diagrams for double photon scattering.
    In the left diagram, the upper, double excited state, $c$ is
    accessed through two photon absorption.  Since this state
    is radiatively unstable, we assume it can decay
    via two photon emission via the intermediate state $b$.  
    On the right, only the intermediate state is accessed
    and the process can be considered to be two independent Raman scattering events. 
    }
    \label{fig:1}
\end{figure*}

Our theoretical approach is to use
the Feynman diagram technique to 
compute the time-integrated two-photon correlation intensity following
either 2-photon scattering or 2-photon radiative cascade from a 
$J$- or $H$-aggregate dimer system.
We assume 
the bi-exciton system can by described by
\begin{eqnarray}
H_{ex} &=& E_x (\sigma_{z1} + \sigma_{z2} + 1)
+ J(\sigma_1^+ \sigma_2^- + \sigma_2^+ \sigma_1^-) \nonumber \\
    &+& U(\sigma_{z1}+1/2)( \sigma_{z2} + 1/2)
\label{eq:1}
\end{eqnarray}
where by the first term corresponds to the uncoupled 
qubits, the second promotes  exchange between qubits
and the third introduces two-body
interactions corresponding to the energy cost to add a second excitation to the 
system.
Writing this in a $SU(2)\otimes SU(2)$ basis 
$H_{ex}$ has 3 excitations above its ground-state
with energies
\begin{subequations}
\begin{eqnarray}
E_{c}  &=& 2E_x + U\,\,\,\,\,\,{\rm biexciton}\\
E_{d}  &=& E_x -J\,\,\,\,\,\,\,\,\,{\rm dark} \label{eq2}\\
E_{b}  &=& E_x + J\,\,\,\,\,\,\,\, {\rm bright} \\
E_{a}  &=& 0\,\,\,\,\,\,\,\,\,\,\,\,\,\,\,\,\,\,\,\,\,\,\,{\rm ground}
\label{eq:2}
\end{eqnarray}
\end{subequations}
Figure~\ref{fig:1}a gives a sketch of relative placement of the energy levels in this system.
In the uncoupled system, an excitation can be placed in either the $|10\rangle$ or $|01\rangle$ state. 
The exchange interaction $J$ splits these states into a symmetric bright state
( $|\psi_b\rangle =
(|10\rangle + |01\rangle)/\sqrt{2}$)
and an anti-symmetric dark state
( $|\psi_d\rangle =
(|10\rangle - |01\rangle)/\sqrt{2}$)
and the anharmonic interaction
shifts the energy of the doubly excited ($|\psi_c\rangle  = |11\rangle$) $U$.
We add to this the radiation field and 
coupling so that the full Hamiltonian becomes
\begin{eqnarray}
H = H_{ex} + H_r + H_I \label{eq:3}
\end{eqnarray}
Under the rotating-wave approximation
$H_I$ would include an exchange term between photons and 
excitons, and hence $|\psi_d\rangle$ becomes ``dark'' and 
all of the photophysics occurs between the ground-state
and states $|\psi_b\rangle$ and $|\psi_c\rangle =|11\rangle$. 
For a $J$-aggregate such as anthracene,
 $J<0$ such that the dark-state $|\psi_d\rangle$ lies above 
the lower $|\psi_b\rangle$ state.  
For $H$-aggregates, the reverse occurs and the ``bright'' state
is higher in energy than the ``dark'' state. The coupling 
term $J$ is mediated by dipole-dipole interactions between transition
densities, the coupling itself depends upon both the 
distance between qubits as well as their relative orientation. 

Since the states are  embedded in the continuum of photon states, they 
acquire an energy shift, $\Delta_n$, and a decay rate $\Gamma_{n}$. Accordingly, we denote such dressed states as 
$| \psi_n\rangle$ and introduce associated complex energy parameter $\tilde E_n  = 
E_n + \Delta_n + i \hbar\Gamma_{n}/2$.  
While the ground-state is also dressed by the radiation field, it is in fact stable, i.e.,  $\Gamma_a = 0$.  
Of the dressed-states, only the ground state is an eigenstate of $H$.
Generally, the energy shifts $\Delta_n$ are very small and can be ignored. 
However, the $\Gamma_n$ should be retained and evaluated from the inverse radiative lifetimes. For $\pi\pi*$ 
transitions in organic materials, radiative lifetime varies in the interval of  
10-100 fs corresponding to the homogeneous linewidth variation of 40-400 meV.  

These assumptions allow us to write the resolvent for a given state $n=\{b,c\}$ as
\begin{eqnarray}
G_n(z) = \langle \psi_n | \frac{1}{z-H} | \psi_n \rangle = \frac{1}{z - E_n + i\hbar  \Gamma_n/2}.
\label{eq:4}
\end{eqnarray}
Also, we introduce a transitions matrix elements between the dressed states as $\mu_{nm}(k)=\langle \psi_n;0 | \hat \mu | \psi_m;k \rangle$ where $k$ is a photon momentum and 0 denotes photon vacuum. The transition dipole operator is determined as $\hat \mu(k) = [H_I,a^\dagger(k)]$. For its the complex (Hermitian) conjugate we use a notation $\mu_{ij}^+(k)$.


Having established the system, we first consider the two photon cascade decay  of the upper most, doubly excited state.
We shall assume that this state can be prepared by an incoherent pump populating the bi-exciton state. Subsequently, the bi-exciton decays via a two-photon cascade that can be detected by a two-photon coincidence measurement.

\subsection{Entanglement by Radiative Cascade}

Before computing the full 2-photon-in$\to$ 2-photon-out scattering process, we first consider the cascaded 
radiative decay from the upper bi-exciton state.   Referring to the left
Feynman diagram in Fig.~\ref{fig:1}b, we prepare the
system in state $c$ and disregard the bottom half of the diagram.
In essence, the two photon process preparing state $c$ can be considered 
as the time-reverse of the 2-photon cascade.  Thus, once we have an expression 
for the cascade, it is trivial to obtain the 2-photon scattering term. 
 Since both $b$ and $c$ are unstable,
they acquire a line-shape and the two photon decay from 
$\tilde \psi_c \to \tilde\psi_0$ only needs to pass through the density 
of states around state $b$.  
We start from state $|\tilde \psi_c;0\rangle$, that is the upper state with no free photons, and decay to the 
ground-state to produce 2 free photons
$|\tilde \psi_0;k_1,k_2\rangle = a^\dagger(k_1) a^\dagger(k_2)|\tilde\psi_0\rangle$.
The amplitude for the $|\tilde \psi_c;0\rangle \to |\tilde \psi_b;k_1 \rangle \to |\tilde \psi_0;k_1k_2\rangle$ transition 
is then 
given by 
\begin{widetext}
\begin{subequations}
\begin{eqnarray}
G(z) &=&
\mu_{0,b}^+(k_2)\mu_{bc}^+(k_1)G_0(z-\hbar\omega_1-\hbar\omega_2)G_b(z-\hbar\omega_1)G_c(z)  \\
&=&
\frac{\mu_{0,b}^+(k_2)\mu_{bc}^+(k_1)}{(z-(\hbar\omega_1 + \hbar\omega_2 + \tilde E_0))(z-\hbar\omega_1 - \tilde E_b)(z-\tilde E_c)}.
\label{eq:5}
\end{eqnarray}
\end{subequations}
Given this, we can calculate the time-dependent amplitude
\begin{eqnarray}
U(\tau) = \lim_{\eta\to 0_+}\frac{1}{2\pi i} \int_{-\infty}^\infty
e^{iE\tau/\hbar}G(E+i\eta)dE.
\label{eq:6}
\end{eqnarray}
From this we get the integrated intensity
$I(\omega_1,\omega_2) = \lim_{\tau \to \infty} |U(\tau)|^2$ as a symmetric 
function of the two frequencies.

Taking $\tau \gg \Gamma_b^{-1}\ \& \ \Gamma_c^{-1} $, only the pole around the ground-state contributes to the integral.
\begin{eqnarray}
U(\tau) = e^{i(E_0 + \hbar\omega_1 + \hbar \omega_2)\tau/\hbar}
\frac{\mu^+_{0b}\mu^+_{bc}}{
(E_0-E_b + \hbar\omega_1 + i \hbar \Gamma_b/2)
(E_0-E_c + \hbar \omega_1+ \hbar \omega_2 +i\hbar\Gamma_c/2)
}.
\label{eq:7}
\end{eqnarray}
Furthermore, we need to include the amplitude corresponding
to the case where the $\omega_1$ photon is emitted after
the $\omega_2$ photon.  Consequently, the 
full amplitude is given by 
\begin{eqnarray}
U_{{\bf k}_1{\bf k}_2}(\tau)=\frac{\mu^+_{0b}\mu^+_{bc}e^{i(E_0 + \hbar\omega_1 + \hbar \omega_2)\tau/\hbar}}{E_o-E_c + \hbar (\omega_1 + \omega_2) + i\hbar\Gamma_c/2}
\left(
\frac{1}{E_o-E_b + \hbar \omega_1 + i\hbar\Gamma_b/2}
+
\frac{1}{E_o-E_b + \hbar \omega_2 + i\hbar\Gamma_b/2}
\right).
\label{eq:8}
\end{eqnarray}
The last two terms can be combined and we write $\hbar\omega_c =E_c-E_o$
and $\hbar\omega_b = E_b-E_o$ and set  $\Gamma_c = 2\Gamma_b = 2\Gamma$.
\begin{eqnarray}
U_{{\bf k}_1{\bf k}_2}(\tau) = 
\frac{\mu^+_{0b}\mu^+_{bc}e^{i(E_0 + \hbar\omega_1 + \hbar \omega_2)\tau/\hbar}}{ \hbar\omega_1 +\hbar \omega_2-\hbar\omega_c + i\hbar\Gamma}
\left(
\frac{\hbar\omega_1 + \hbar\omega_2 - 2 \hbar\omega_b + i \hbar\Gamma}
{(\hbar\omega_1-\hbar\omega_b + i\hbar \Gamma/2)
(\hbar\omega_2-\hbar\omega_b + i\hbar \Gamma/2)
}
\right).
\label{eq:9}
\end{eqnarray}
Hence, one can write the entangled intensity as a normalized 
probability distribution
\begin{eqnarray}
I(\omega_1,\omega_2) &=& \lim_{\tau\to\infty}\left|U(\tau)\right|^2
\label{eq:10}
\end{eqnarray}

We here consider a special case where $\omega_c = 2\omega_b$, 
corresponding to the case that the intermediate level is 
exactly half way between level $c$ and the ground state.
For our model system, this occurs when 
$J = U/2$ . 
For a $J$-aggregate in which $J<0$ and $U>0$ the condition that $2\omega_b = \omega_c$ can not be satisfied.  
For an $H$-aggregate, however, this condition can be satisfied over a range of both $J$ and $U$.
Under this special case condition
Eq.~\ref{eq:9} can be 
further factored to
\begin{eqnarray}
U_{{\bf k}_1{\bf k}_2}(\tau) =
e^{i(E_0 + \hbar\omega_1 + \hbar \omega_2)\tau/\hbar}
\left(
\frac{\mu^+_{0b}\mu^+_{bc}}
{(\hbar\omega_1-\hbar\omega_b + i\hbar \Gamma/2)
(\hbar\omega_2-\hbar\omega_b + i\hbar \Gamma/2)
}
\right).
\label{eq:11}
\end{eqnarray}
In this case, the integrated intensity can by written as 
a purely separable function of the two frequencies.
\begin{eqnarray}
I(\omega_1,\omega_2) = \frac{|\mu_{0b}|^2|\mu_{bc}|^2}{\hbar^4}
\frac{1}{(\omega_1-\omega_b)^2 + \Gamma^2/4}
\frac{1}{(\omega_2-\omega_b)^2 + \Gamma^2/4}.
\label{eq:12}
\end{eqnarray}

\end{widetext}

\begin{figure*}
    \centering
    \includegraphics[width=0.6\columnwidth]{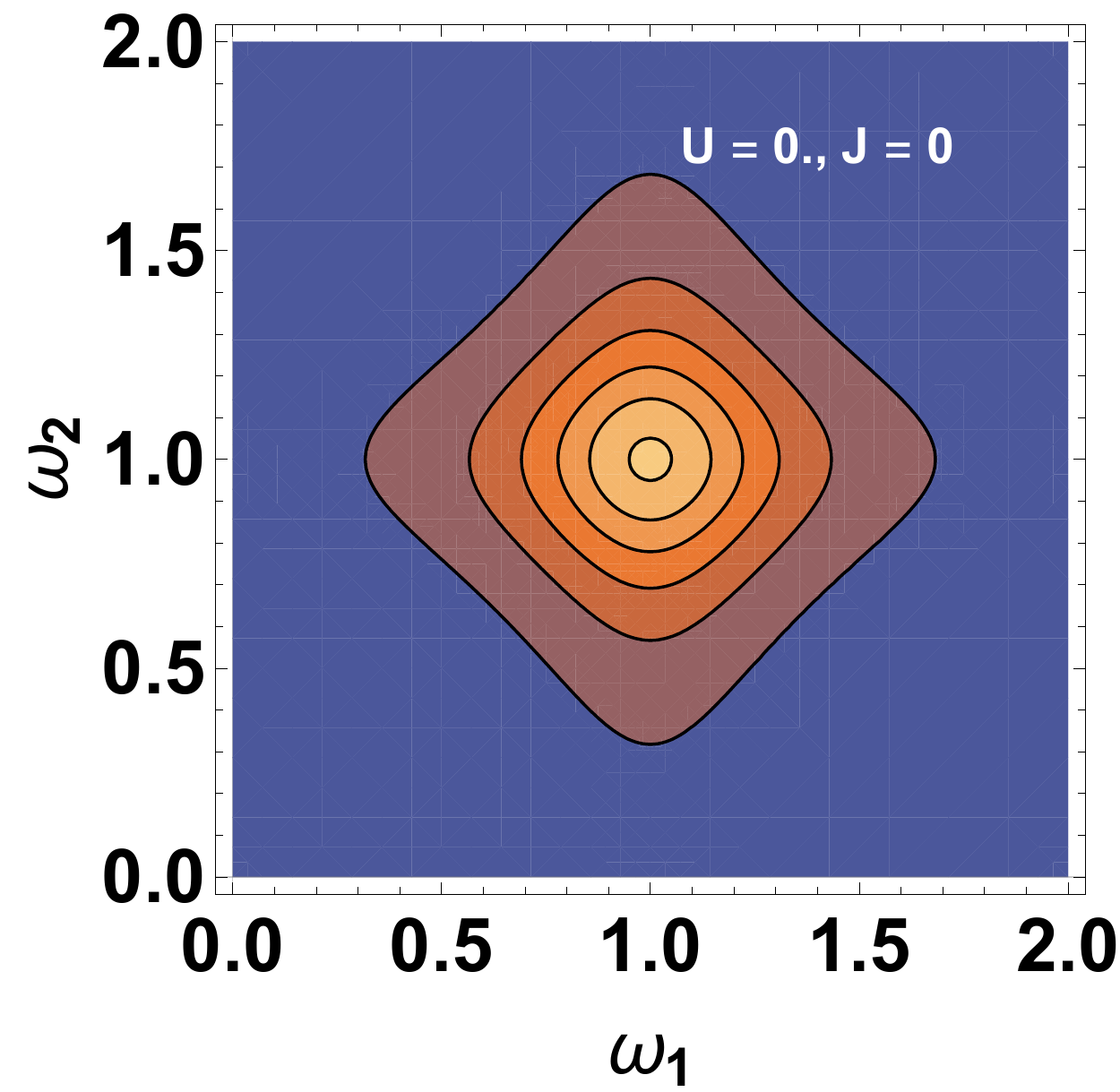}
    \includegraphics[width=0.6\columnwidth]{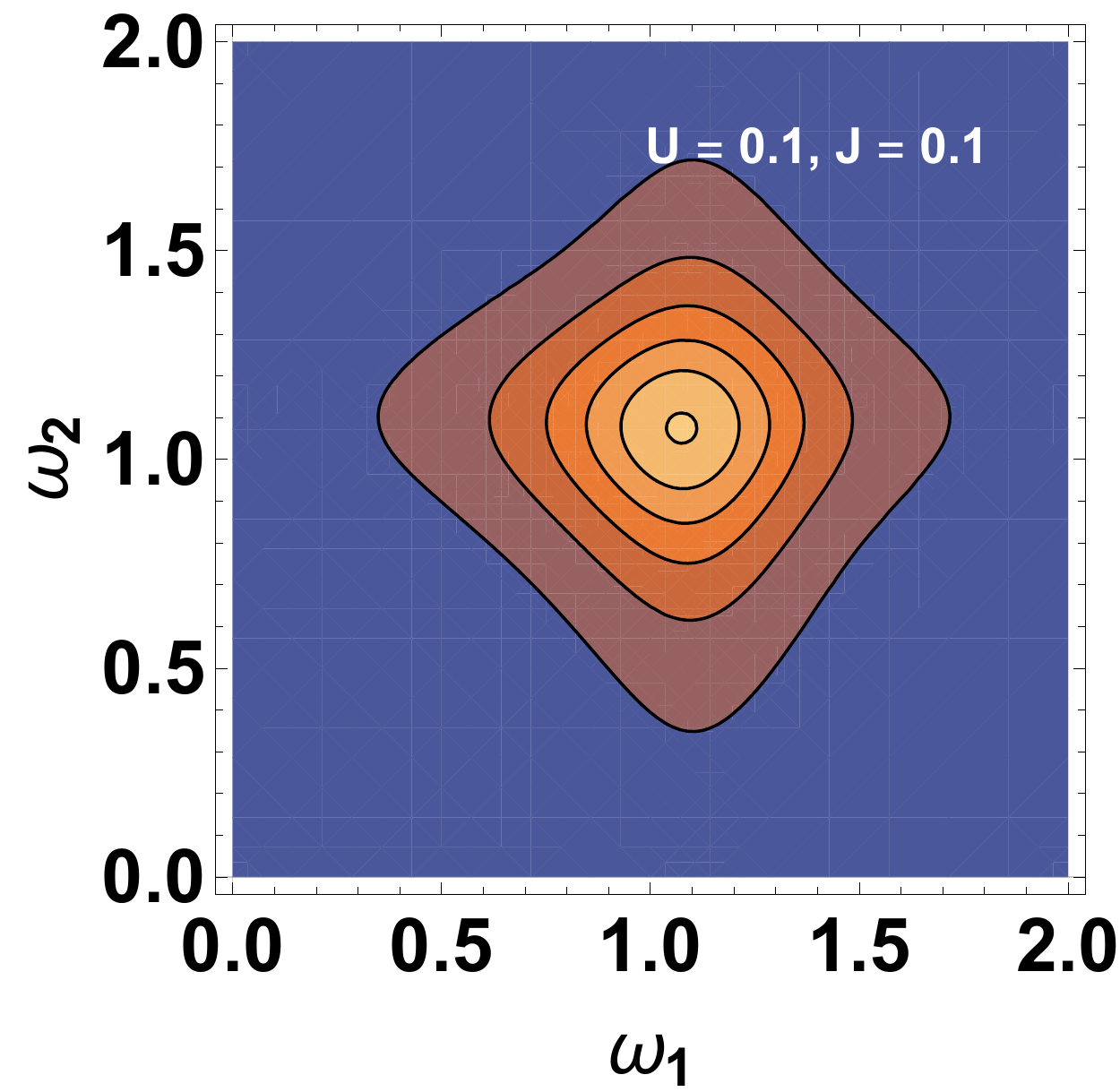}  
    \includegraphics[width=0.6\columnwidth]{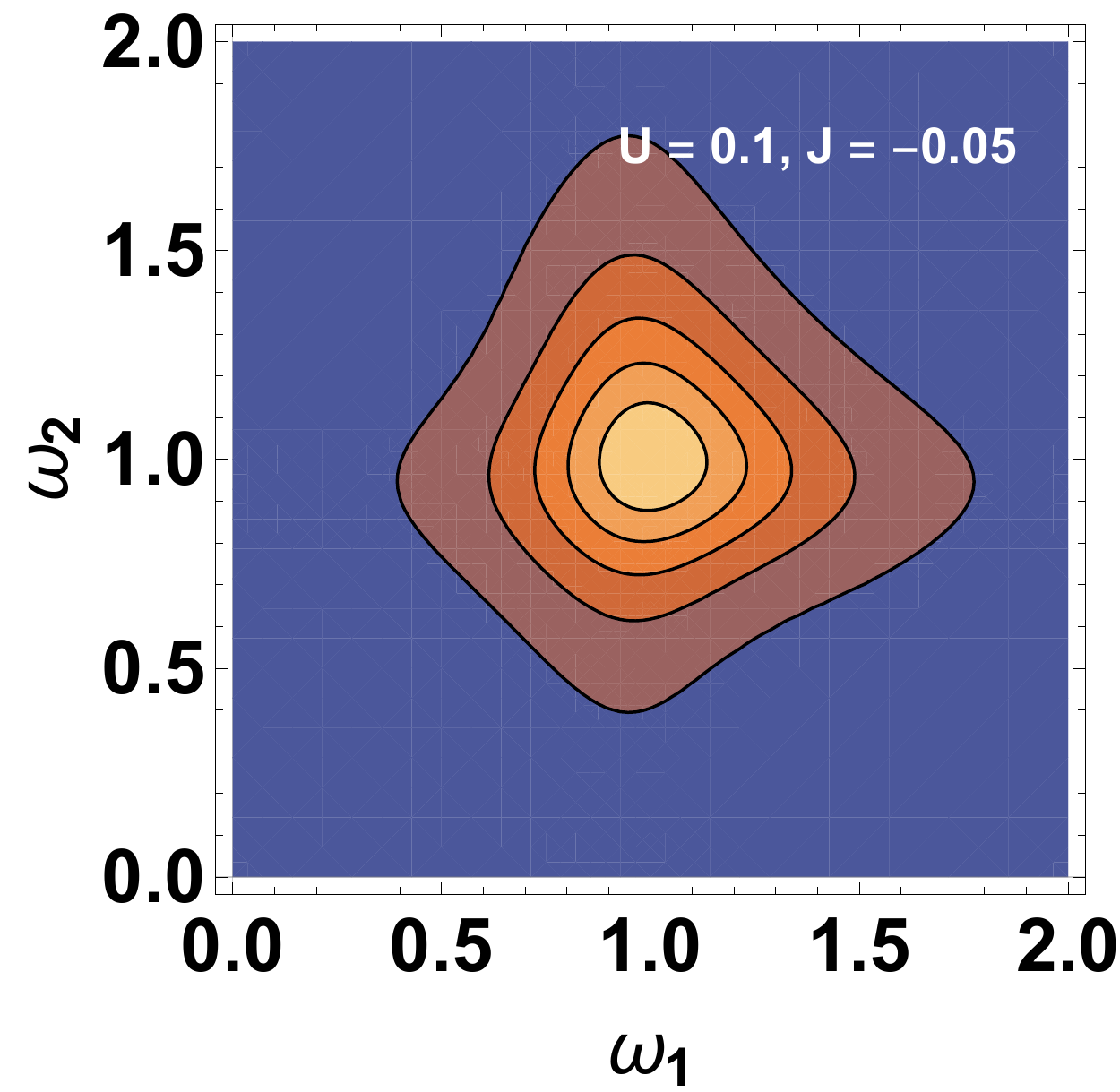}
    \caption{Two-photon cascade emission 
    intensity distributions as varied by 
    exciton/exciton interaction terms, $U$, and $J$. (a) Non-interacting qubits. (b) $H$-aggregate ($U>0$ \& $J>0$). (c) $J$-aggregate ($U>0$ \& $J<0$).}
    \label{fig:2}
\end{figure*}

In Figure~\ref{fig:2}(a-c) we show the integrated 
intensity $I(\omega_1,\omega_2)$ for 
two-photon emission from a bi-exciton state in a model $J$-aggregate
dimer system with $\hbar\omega_x = 1$ setting the 
energy scale and with $U$ and $J$ indicated
on the plots.  In each, we assume $\Gamma = 0.6$ which 
in consistent with a lifetime of about 
6fs for an exciton with $\hbar\omega_x = 1eV$. 
The line shape is symmetric
about the line $\omega_1 = \omega_2$ reflecting the 
fact that we summed over both photon paths. 
In the uncoupled  case, $U=0$ and $J=0$, the distribution 
is clearly separable into two terms. 
Increasing the either the
hopping term $J$ or the repulsion term $U$ leads to intensity 
distributions that are no longer separable into single photon terms.

\subsection{Double photon scattering}
We now consider the case depicted in the 
left-most Feynman diagram in Fig.~\ref{fig:1}b.
In this case, the two input photons place the system into the 
doubly-excited $|c\rangle = |11\rangle$
state and are re-emitted leaving the 
system back in is ground state:
$|a\rangle \to |b\rangle \to |c\rangle \to |b'\rangle \to | a'\rangle$.

For this, we shall write the amplitude in terms of the 
M{\o}ller operators to propagate the initial state $|a,n_1,n_2\rangle$
from $t\to -\infty$ forward to some intermediate time $t$, 
where the system is in the $|c,n_1-1,n_2-1\rangle$ state,
then from $t\to +\infty$ to $|a,n'_1,n'_2\rangle$. 
The M{\o}ller operator interwines the asymptotic (i.e. free) observables
to those in the fully interacting theory.  These are 
especially important considering the scattering of quantum photons
since the atomic/material target is never fully free of the 
radiation field.\cite{Atom-photon-interactions} 
The operators are defined by 
writing the interaction picture ket as
\begin{eqnarray}
    |\psi(t)\rangle_I = e^{iH_{o}t/\hbar}e^{-i(H_{o}+V)t/\hbar}
    |\psi\rangle
    \label{eq:13}
\end{eqnarray}
where $|\psi\rangle$ is the asymptotic state. Inverting this, 
\begin{eqnarray}
    |\psi\rangle=e^{i(H_{o}+V)t/\hbar} e^{-iH_{o}t/\hbar}
    |\psi(t)\rangle_I.
     \label{eq:14}
\end{eqnarray} 
Upon taking the limits of $t\to \pm\infty$, one defines the M\o ller operators\cite{MOLLER:1946aa,MollerOriginal1945}
\begin{eqnarray}
    \Omega^{(\pm)}=\lim_{t\to \mp\infty}e^{i(H_{o}+V)t/\hbar} e^{-iH_{o}t/\hbar}.
      \label{eq:15}
\end{eqnarray} 
Assuming we have two photons in the asymptotic states, the
relevant states are
\begin{subequations}
\begin{eqnarray}
    |\psi^{-}({\bf k}_1,{\bf k}_2) \rangle &=& a_{{\bf k}_1}^\dagger a_{{\bf k}_2}^\dagger |a;0\rangle \\
    |\psi^{+}({\bf k}_1',{\bf k}_2') \rangle &=& a_{{\bf k}'_1}^\dagger a_{{\bf k}'_2}^\dagger |a';0\rangle.
    \label{eq:16}
\end{eqnarray}
\end{subequations}
Thus we transform the input state $ |\psi^{-}({\bf k}_1,{\bf k}_2) \rangle$ to the output state
\begin{eqnarray}
|\psi^+({\bf k}_1',{\bf k}_2')\rangle &=& \Omega^{(-)\dagger}\Omega^{(+)}|\psi^-({\bf k}_1,{\bf k}_2)\rangle \\
&=& \hat S^{(2)}|\psi^-({\bf k}_1,{\bf k}_2)\rangle
\label{eq:18}
\end{eqnarray}
where $S$ is the scattering matrix. 
Since the initial and final atomic states will be the same ground state,$a = a'$,  energy and momentum conservation will require that
$\hbar\omega_1 + \hbar\omega_2 = \hbar\omega_1' + \hbar\omega_2'$ and ${\bf k}_1 + {\bf k}_2 = {\bf k}'_1 + {\bf k}'_2$.
The final transition amplitude can now be deduced from 
Eq.~\ref{eq:9} by forward propagating the input state and reverse propagating
the final state to some intermediate time $\tau$ where the system is
in $|c\rangle$.  
\begin{eqnarray}
    S^{(2)} = \iint_{-\infty}^{+\infty} d \tau d \tau' U_{k_1k_2}^\dagger(\tau) U_{k'_1k'_2}(\tau')
\end{eqnarray}
Integrating over all intermediate times
 the energy conservation
$\omega_1 + \omega_2 = \omega_1' + \omega_2'$,
and finally one finds that
\begin{widetext}

\begin{eqnarray}
\hat S^{(2)}(\omega_1,\omega_2;\omega_1',\omega_2') &=&  
\frac{\mu_{0b}\mu_{bc}\mu^+_{0b}\mu^+_{bc}}{\hbar^4} 
\left(
\frac{(\omega_1 + \omega_2 - 2 \omega_b)^2 +  \Gamma^2}
{ (\omega_1 + \omega_2-\omega_c)^2 + \Gamma^2}
\right)\nonumber \\
&\times&
\left(
\frac{1}
{(\omega_1-\omega_b - i \Gamma/2)
(\omega_2-\omega_b - i \Gamma/2)
}
\frac{1}
{(\omega'_1-\omega_b + i \Gamma/2)
(\omega'_2-\omega_b + i \Gamma/2)
}
\right).\label{2photonSmat}
\end{eqnarray}
\end{widetext}
Since the general form of the two-photon scattering 
matrix is identical in form to what we arrived at for the
cascade (aside from a constant term), the resulting 
entanglement change reflects the cascade dynamics from 
state $|c\rangle$.

\begin{figure*}
    \centering
	{\includegraphics[width=0.6\columnwidth]{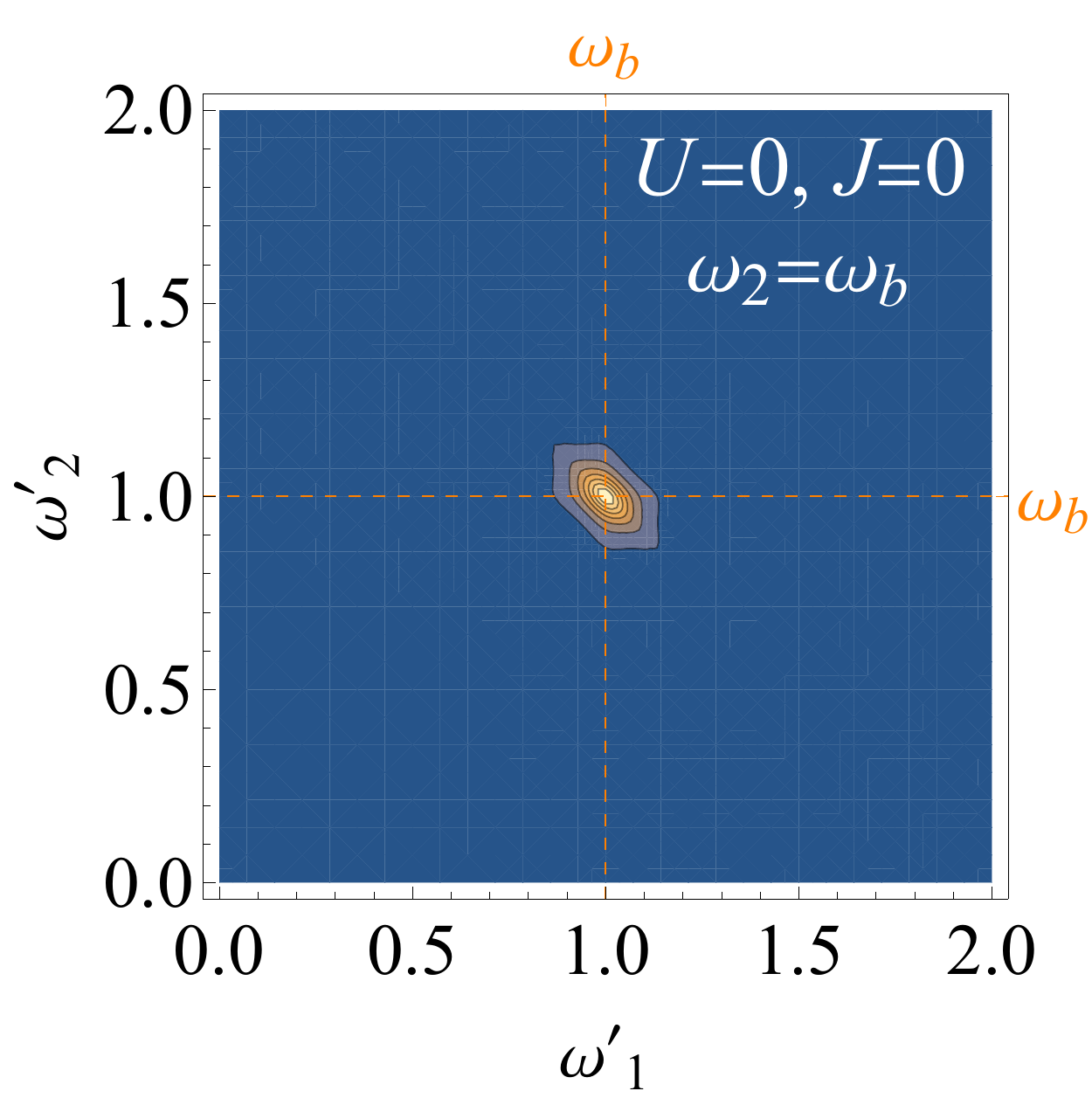}}
	{\includegraphics[width=0.6\columnwidth]{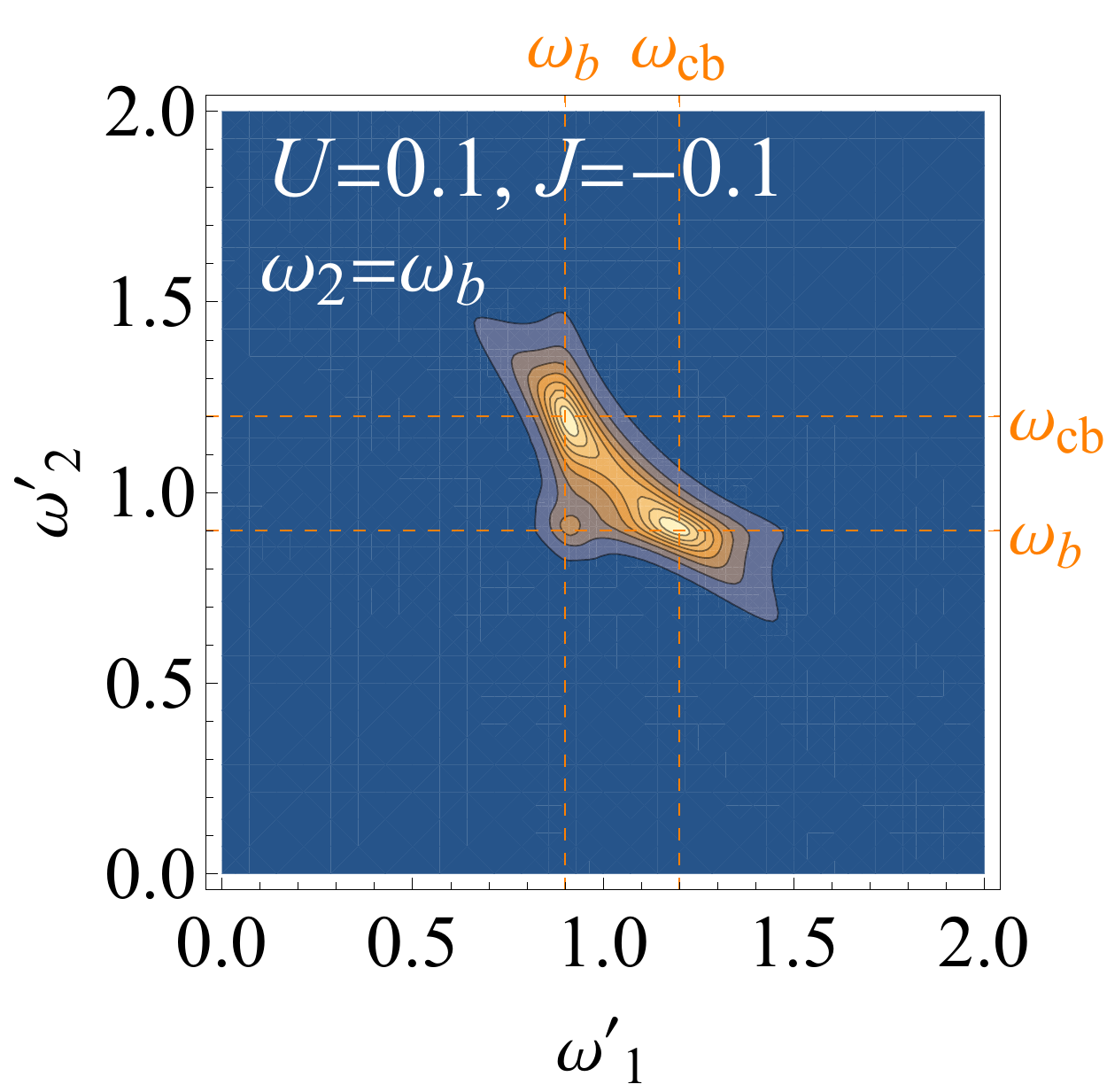}}
	{\includegraphics[width=0.6\columnwidth]{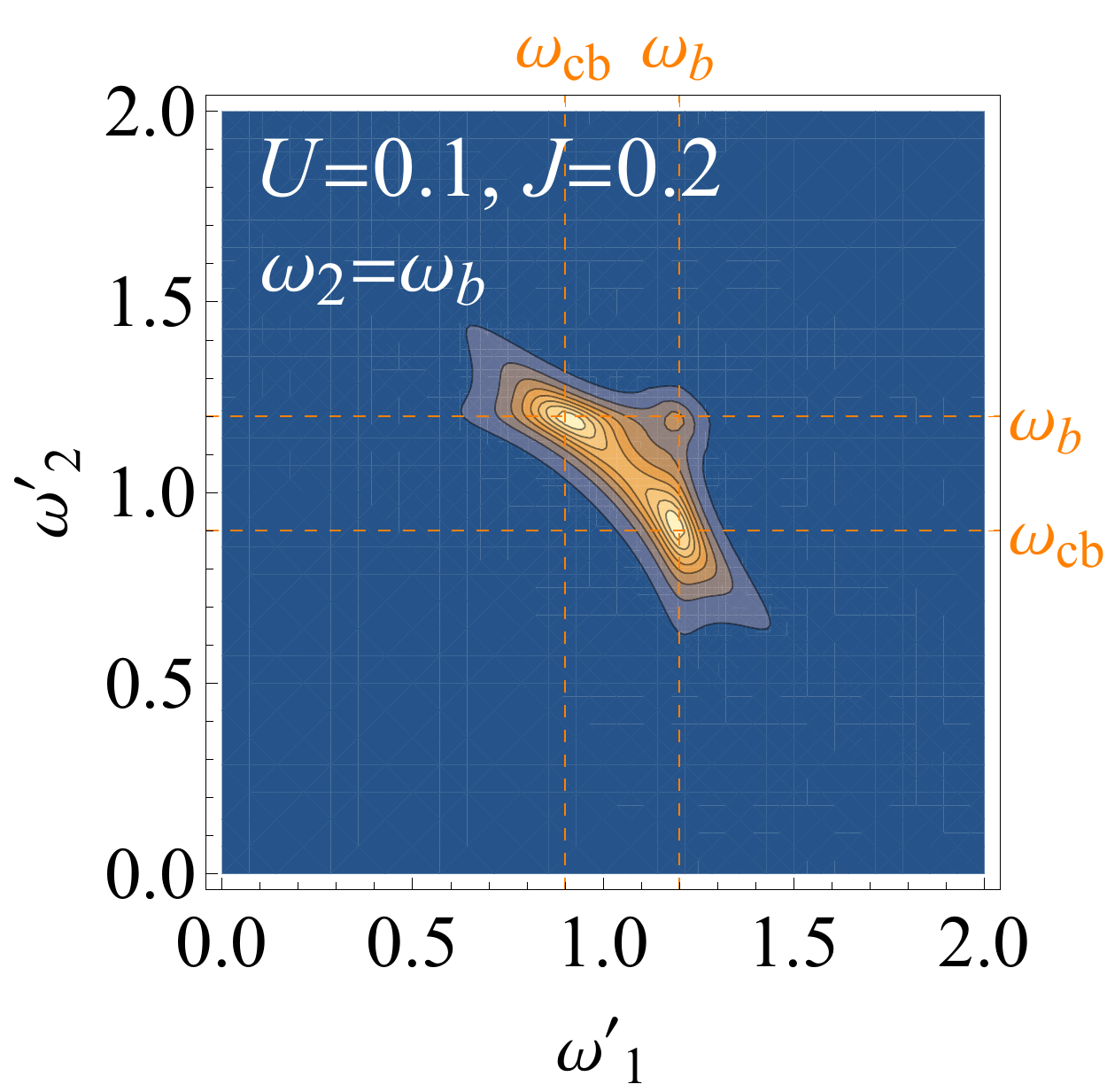}}
	{\includegraphics[width=0.6\columnwidth]{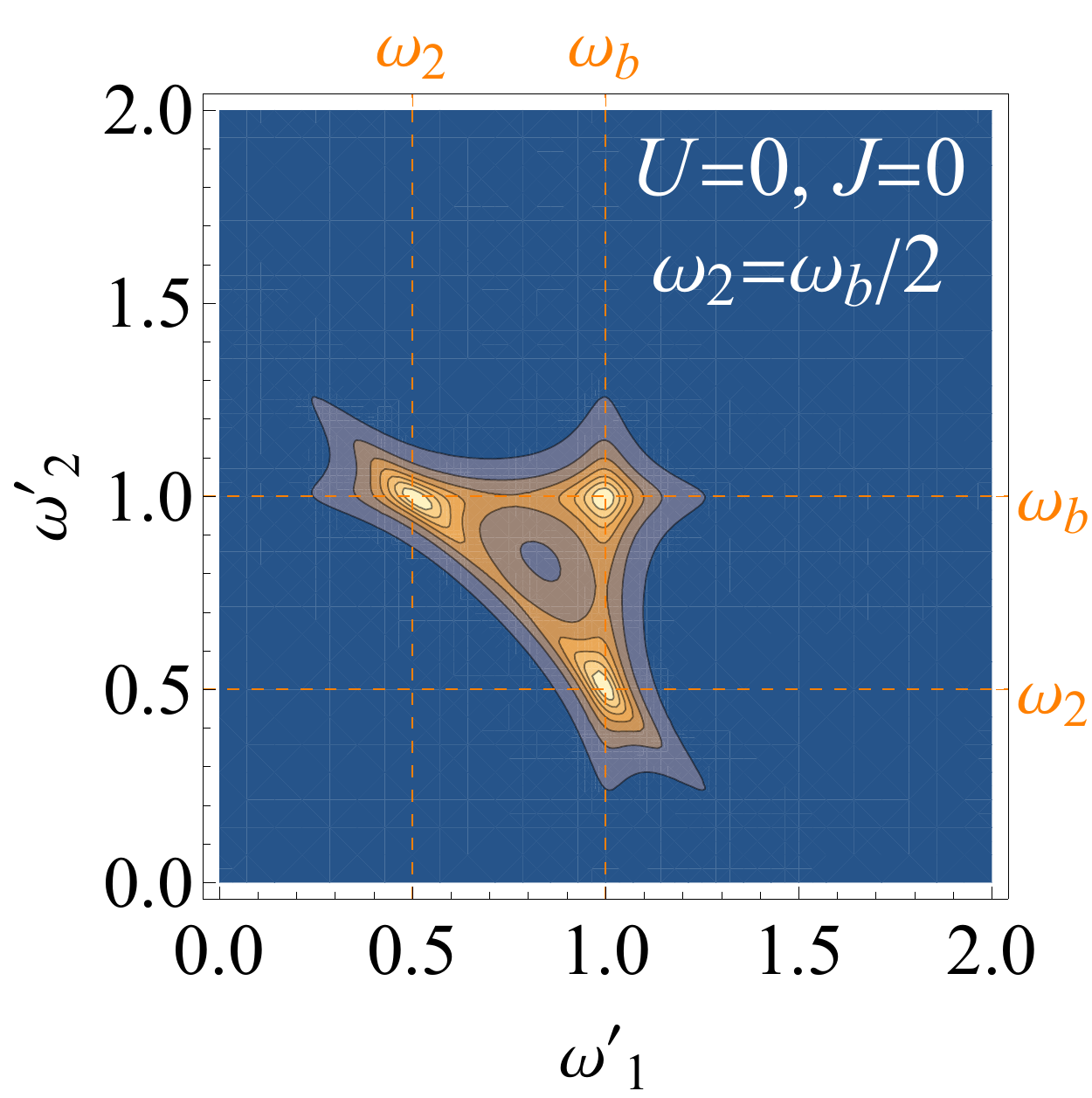}}
	{\includegraphics[width=0.6\columnwidth]{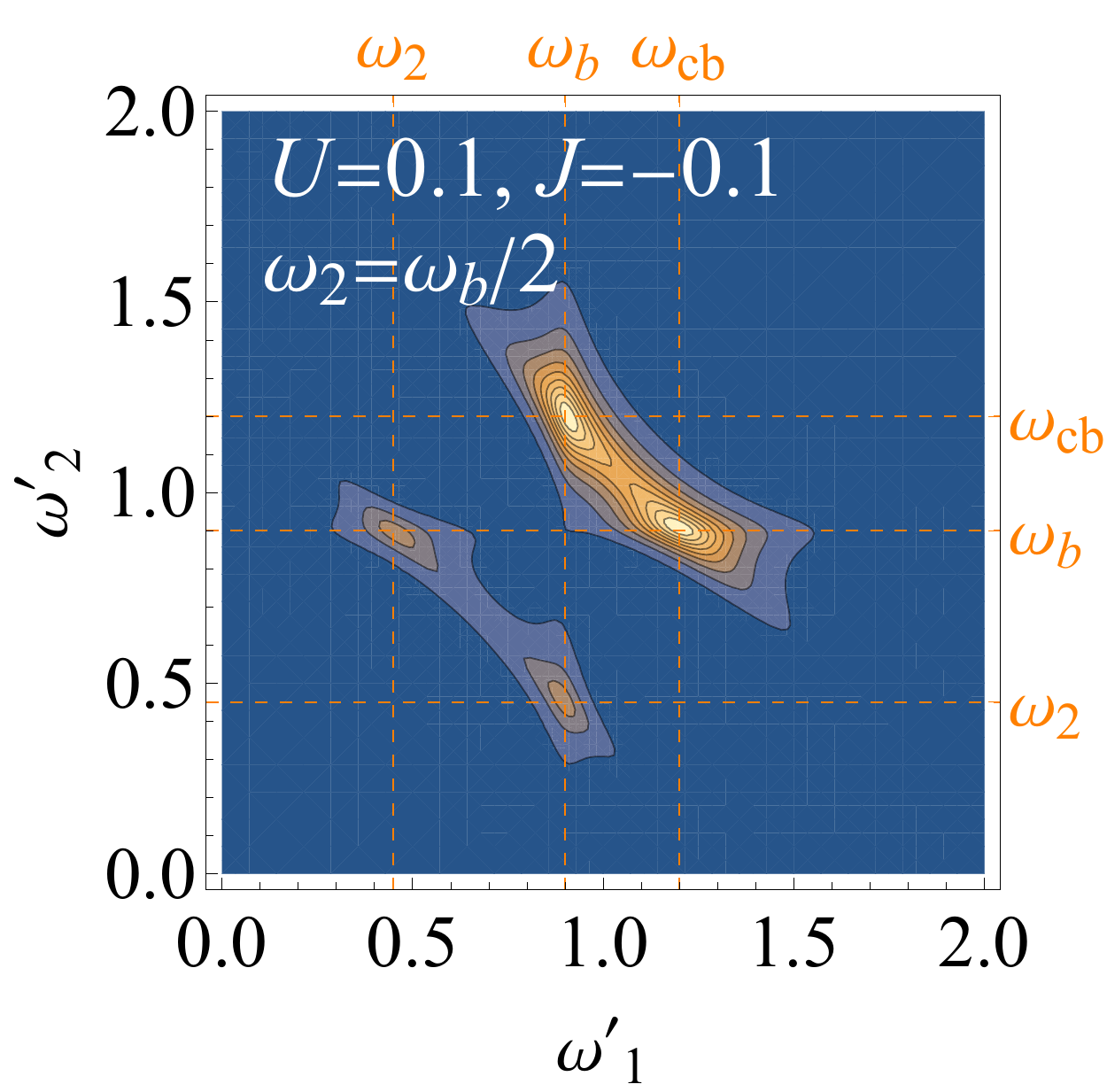}}
	{\includegraphics[width=0.6\columnwidth]{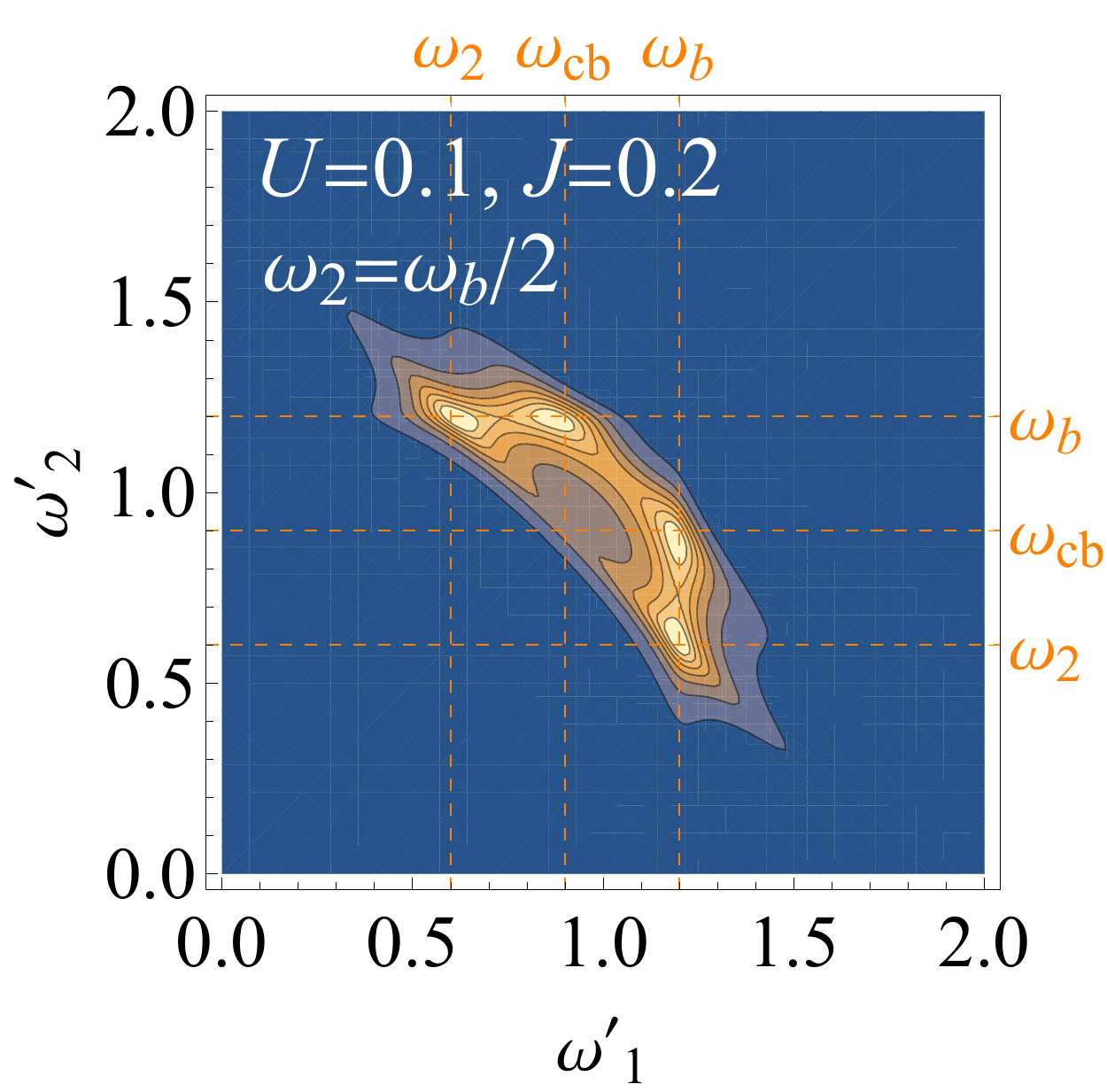}}
	\caption{Two-photon scattering distributions in terms of output photon frequencies for non-interacting dimer (left column), $J$ and $H$ aggregate systems (center and right columns, respectively). One of the input photon energy $\omega_1$ is free to change whereas the other is set to be either on-resonant ($\omega_2 =\omega_b$, upper panels) or off-resonant (($\omega_2 =\omega_b/2$, lower panels) with the bright state $|b \rangle$. Note that we set $\Gamma=0.1E_x/\hbar$ and $\omega_{bc}=\omega_c-\omega_b$ for all plots.}
	\label{fig:2dscat}
\end{figure*}

In Fig.~\ref{fig:2dscat} we show the results for scattering 
an initial input bi-photon Fock state $|\omega_1\omega_2\rangle$ 
under various parametric conditions. The output bi-photon state is correlated to the input due to the energy and momentum conservation. Therefore, we allow the frequency of one input photon to vary freely but fix the other to be either resonant with the bright state $|b \rangle$, or very much off-resonant by setting $\omega_2=\omega_b/2$. The value of the
exciton transfer term $J$ is deliberately chosen to be large to highlight
the effect of resonant coupling between the two qubits. 
Clearly, dimer interaction has a profound effect upon the scattering behavior. No matter one of the input photons is on-resonance or off-resonance, distinguished scattering can always be observed around $\omega_1+\omega_2=\omega_c$, i.e., the input bi-photon state being resonant with the bi-exciton state, whereas the scattering is weak in the case of single photon resonance $\omega_1=\omega_b$. The outgoing
state is always entangled due to the initial entanglement and the photon-photon coupling introduced by 
interactions with the medium.~\cite{Kalashnikov2016a,Yabushita:PRA2004}

\subsection{Entanglement Entropy Generation}

The entropy $S$ provides a useful metric for the 
entanglement carried by the outgoing photons. This can be determined by 
singular value decomposition of $I(\omega_1,\omega_2)$ in which 
we write the 2-photon intensity
\begin{eqnarray}
I(\omega_1,\omega_2) = \sum_n r_n f_n(\omega_1) g_n(\omega_2)\label{eq16}
\end{eqnarray}
as a weighted sum over single-component terms determined by Singular Value
Decomposition (SVD). Taking $r_n$ to be
normalized to unity, 
\begin{eqnarray}
S = -\sum_n r_n \ln r_n.\label{eq17}
\end{eqnarray}
The functions $f_n(z)$ and $g_n(z)$ are orthogonal polynomials
forming a complete basis. 

\begin{figure}[t]
	\centering
	\includegraphics[width=\columnwidth]{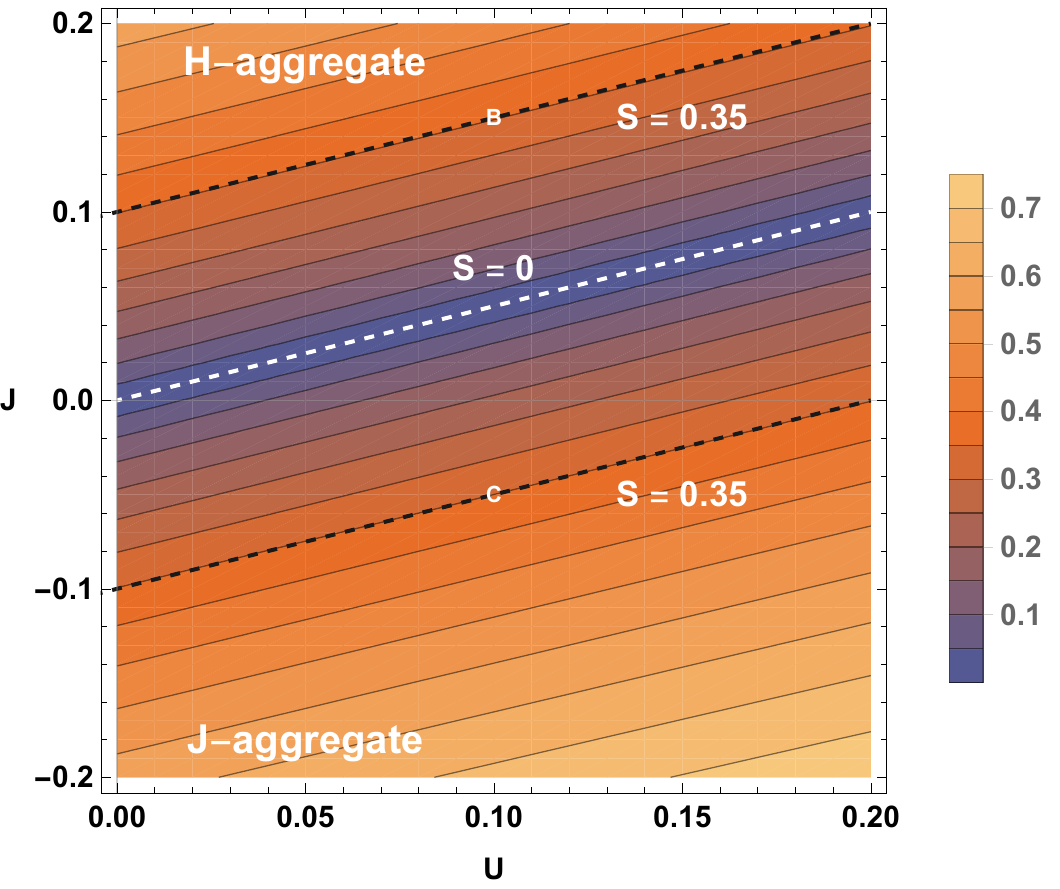}
	\caption{Entanglement entropy for model interacting binary qubit system as
		varied by the interaction parameters $U$ and $J$. Units are 
		such that $E_x = 1$.
		The white line corresponds to the condition where $2J = U $ whereby 
		the single excitation energy is exactly half the double-excitation 
		energy.
	}
	\label{fig:3}
\end{figure}

In Fig.~\ref{fig:3} we computed the entropy as function of both $J$ and $U$ 
over a wide parametric range.  
The upper half ($J>0;U>0$) corresponds to the situation for most $H$-aggregate systems. 
Here, the positions of the two middle energy levels are swapped and the (now) upper state is what 
carries the coupling to the radiation field.  
In this regime, we have the possibility for 
satisfying the $2\omega_b = \omega_c$ criterion for a fully separable two-photon emission spectrum. 
The lower half ($J<0; U>0$) is corresponds to the 
parametric range for $J$-aggregate systems.
Here, because the $2\omega_b = \omega_c$ cannot be
satisfied, the entanglements are higher.
For comparison, we consider two systems with 
identical entanglement entropy indicated on Fig.~\ref{fig:3} by the letters ``B" and ``C"
corresponding to an $H$-aggregate 
(Fig.~\ref{fig:2}b) and
a $J$-aggregate (Fig.~\ref{fig:2}c).
Since the coupling terms are sensitive to 
packing and aggregation, it is should be possible to
control and select the entanglement in the emitted 
photon state.

The right-most Feynman diagram 
in Fig.~\ref{fig:1}b corresponds to 
two successive Raman scattering processes.
In this case, the two processes will be independent 
in the limit that the line-shape of state $b$ is sufficiently
broad.\cite{Li:2019} 
It is important
to distinguish this process from the double-excitation
process discussed above.
Assuming the two $a\to b \to a'$ process are uncorrelated, 
the integrated intensity is the product of individual 
Raman intensities. 
As we showed in our previous work, even if 
the second excitation occurs within the 
homogeneous lifetime $\Gamma_b^{-1}$ of state $b$ the 
two events will not produce entangled photons. 
However, in the limit of slow-modulation
the two transition moments can be correlated 
giving rise to entanglement in the outgoing photon state.

\section{Discussion}

In this paper we have shown how to construct the transition matrix for 2-photon 
resonant scattering which produces entanglements within the out-going photon state.  We show that 
such an entanglement can be connected to 
exciton/exciton interactions occurring via exchange 
coupling between sites and exciton/exciton repulsion
within a binary qubit system that corresponds to 
a molecular dimer. 
Since these parameters are exquisitely linked to the 
local structure of the system and relative 
orientation of the transition dipoles on each 
 monomer, it should be possible to 
manipulate the outgoing entanglement via external
means.   

While we present results for a simple
excitonic dimer, the model and methods
are easily extendable to systems with multiple 
excitonic sites and 
internal vibronic degrees of freedom.    Experiments based upon these ideas may 
offer valuable insights into the correlated dynamics 
occurring within complex excitonic systems.
Our current results are valid only in the limit of 
low temperature and where vibronic coupling can be ignored. 
The inclusion of finite temperature and vibronic dynamics
will certainly muddle the waters by limiting the
time-frame over which entanglements can be established.
As discussed in our recent paper, entanglement between the 
outgoing photons is contingent upon the both strength of the interaction
and the magnitude of environmental fluctuations. 
\cite{Li:2019}
Our current efforts are to include both implicit and explicit
quantized vibronic modes into the bi-photon scattering model.

\acknowledgements

The work at the University of Houston was funded in
part by the  National Science Foundation (
MRI-1531814, 
CHE-1664971, 
CHE-1836080  
)
and the Robert A. Welch Foundation (E-1337).
ERB also acknowledges the Leverhulme Trust for 
support at Durham University. 
AP acknowledges the support provided by Los Alamos National Laboratory Directed Research and Development (LDRD) Funds.
CS acknowledges support from the School of Chemistry \& Biochemistry and the College of Science of Georgia Tech. The work in Georgia Tech is supported by the National Science Foundation (CHE-1836075). 
ARSK acknowledges funding from  EU Horizon 2020 via Marie Sklodowska Curie Fellowship (Global) (Project No. 705874).

\bibliography{References-local}
\end{document}